# Follow-up observations for the Asteroid Catalog using AKARI Spectroscopic Observations

Sunao HASEGAWA,[1,*] Daisuke KURODA,[2] Kenshi YANAGISAWA,[2] and Fumihiko USUI[3]

[1]Institute of Space and Astronautical Science, Japan Aerospace Exploration Agency, 3-1-1 Yoshinodai, Chuo-ku, Sagamihara, Kanagawa 252-5210, Japan

[2]Okayama Astrophysical Observatory, National Astronomical Observatory of Japan, 3037-5 Honjo, Kamogata-cho, Asakuchi, Okayama 719-0232, Japan

[3]Center for Planetary Science, Graduate School of Science, Kobe University, 7-1-48, Minatojimaminamimachi, Chuo-ku, Kobe, Hyogo 650-0047, Japan

[*]E-mail: hasehase@isas.jaxa.jp

## Abstract

In the 1–2.5 $\mu$m range, spectroscopic observations are made on the AcuA-spec asteroids, whose spectra were obtained in a continuous covered mode between 2.5–5.0 $\mu$m by AKARI. Based on the Bus-DeMeo taxonomy (DeMeo et al. 2009, Icarus, 202, 160), all the AcuA-spec asteroids are classified, using the published and our observational data. Additionally, taking advantage of the Bus-DeMeo taxonomy characteristics, we constrain the characteristic each spectral type by combining the taxonomy results with the other physical observational data from colorimetry, polarimetry, radar, and radiometry. As a result, it is suggested that certain C-, Cb-, B-type, dark X-, and D-complex asteroids have spectral properties compatible with those of anhydrous interplanetary dust particles with tiny bright material, such as water ice. This supports the proposal regarding the C-complex asteroids (Vernazza et al. 2015, ApJ, 806, 204; 2017, AJ, 153, 72). A combination of the Bus-DeMeo taxonomy for AcuA-spec asteroids and the presumptions with other physical clues such as the polarimetric inversion angle, radar albedo, and mid-infrared spectroscopic spectra will be beneficial for surface material constraints, from the AcuA-spec asteroid observations.

## 1 Introduction

Spectroscopic observation is a powerful method for obtaining surface information on asteroids. Light emitted from asteroids between the ultraviolet and near-infrared wavelength region is superior in reflecting sun light, than the thermal radiation of the asteroids; hence, information on the asteroid surface layer composition can be obtained from their reflected spectra. Olivine and pyroxene, important minerals in meteorites, have characteristic absorption bands in the near-infrared wavelength region of the wavelength range from 0.7– 2.5 $\mu$m (e.g., Reddy et al. 2015); a strong absorption of the hydrated minerals and water ice exists over the wavelength range, 2.6–3.2 $\mu$m (e.g., Rivkin et al. 2015).

In ground observations, certain wavelength regions cannot be observed due to low atmospheric transmission. In particular, observation over wavelengths between 2.5–2.85 $\mu$m with the characteristic absorption of hydrated minerals, is difficult owing to deep atmospheric absorption (e.g., Rivkin et al. 2015). For observation, without the influence of the Earth's atmosphere, it is necessary to observe from space.

AKARI, the first Japanese telescope dedicated to infrared astronomy, was designed as an all-sky survey mission in the mid- and far-infrared wavelength regions (Murakami et al. 2007). It also has the capability to make pointed observations from the near- to far- infrared wavelength regions. AKARI has detected approximately five thousand asteroids in the survey mode (Usui et al. 2011) and photometric asteroid observations were performed in the pointed observation mode (Hasegawa et al. 2008, 2013; Müller, Hasegawa, & Usui 2014). Seamless spectroscopic observations, including the 2.5–2.85 $\mu$m wavelength region, were performed and the first spectroscopic survey catalogue of 65 main-belt asteroids included in the Asteroid Catalog using AKARI Spectroscopic Observations (AcuA-spec), in the wavelength range of 2.5–5.0 $\mu$m, were made (Usui et al. in prep.). AcuA-spec asteroid candidates were selected from the main-belt asteroids only, classified based on the visible spectrophotometric and spectroscopic observations (Tholen 1994; Bus & Binzel 2002b; Lazzaro et al. 2004). Approximately 200 classified asteroids were considered as observational candidates. Spectroscopic observations for the 65 AcuA-spec asteroids, which were selected from the AcuA-spec candidate catalogue considering the signal-to-noise ratio and the target opportunity by AKARI, were performed.

The selections for the AcuA-spec candidate catalogue were made around 2005, yet there were not many spectroscopic asteroid observations in the near-infrared wavelength range. However, of late, several excellent infrared observation devices such as the SpeX instrument (Rayner et al. 2003) are available, owing to the developments in near-infrared technology. Consequently, the spectroscopic observations of numerous asteroids in the near-infrared region have been carried out. From a survey in a published paper regarding the near-infrared spectroscopic observations of asteroids, it was found that spectroscopic observations in the near-infrared region of 1–2.5 $\mu$m for most AcuA-spec asteroids, have been performed, except for certain asteroids. If data in the near-infrared wavelength range for these AcuA-spec asteroids are acquired, then data for all the AcuA-spec asteroids in the 0.35–5 $\mu$m range, will be available[1]. The continuous reflectance of asteroids is essential data for comparison with those of meteorites. The first objective of this study is to obtain spectroscopic data spanning the 1–2.5 $\mu$m wavelength region for all the AcuA-spec asteroids.

Taxonomy is a useful tool for classifying asteroids based on their reflectance spectra. Asteroid taxonomy commenced in the 1970's (Chapman, Morrison, & Zellner 1975) and asteroids were classified into three types. Tholen (1994) classified 14 spectral types, using the wavelength data of eight selected narrow band filters in the visible wavelength region for approximately 400 asteroids by principal component analysis (PCA). Subsequently, Bus & Binzel (2002b) conducted a new taxonomy using the spectroscopic data in the visible wavelength region, on approximately 1500 asteroids, by PCA. After the selection of the AcuA-spec asteroids, new taxonomic methods including the spectroscopic results of the near-infrared wavelength regions up to 2.5 $\mu$m have emerged, along with the recent increase in the near-infrared asteroid spectra. DeMeo et al. (2009) presented the Bus-DeMeo taxonomy, using both visible and near-infrared data to group asteroids spectrally from 0.45–2.45 $\mu$m according to the PCA. The Bus-DeMeo taxonomy is considered to be the most reliable classification for asteroids, covering the characteristic absorption bands of olivine and pyroxene, and the spectral slope over a wide wavelength range. Therefore, the second objective of this study is to classify all the AcuA-spec asteroids, based on the Bus-DeMeo taxonomy.

After the publication of the Bus-DeMeo taxonomy in 2009, numerous asteroids were classified (Clark et al. 2009; Clark et al. 2010; Fornasier et al. 2010; Ockert-Bell et al. 2010; Fornasier, Clark, & Dotto 2011; Ostrowski et al. 2011; de León et al. 2012; Gietzen et al. 2012; DeMeo, Binzel, & Lockhart 2014; Neeley et al. 2014; Polishook et al. 2014), using the taxonomy. As the Bus-DeMeo taxonomy provides excellent classification, the classification may further constrain the physical properties of asteroids by combining with data, not used in the taxonomy. Thus, the third objective of this study is the extraction of information on the physical properties of each spectral type asteroid and to ascertain the properties of the AcuA-spec asteroids.

---

[1] All the UBV colors of the AcuA-spec asteroids have already been published (Tedesco 2005; Toth 1997).

As the first step towards this objective, the classified asteroids in the Bus-DeMeo taxonomy were sought and the corresponding data were compiled. The second step is to combine the compiled Bus-DeMeo taxonomic data with other observation data. The third step is to determine asteroids that corresponds to each extraterrestrial material sample, such as meteorites. As the final step, these considerations are adapted for the AcuA-spec asteroids.

## 2 Observations and data reduction procedures for AcuA-spec asteroids

Near-infrared spectroscopic data spanning the wavelength region between 1–2.5 $\mu$m for 94 Aurora, 127 Johanna, and 423 Diotima was not published, before this study was planned [2]. The broad-band photometric data of 127 Johann and 185 Eunike in the near-infrared wavelength regions was recorded by the deep European near-infrared southern sky (Baudrand, Bec-Borsenberger, & Borsenberger 2004) and the two-micron all-sky surveys (Sykes et al. 2010). However, these data were insufficient for Bus-DeMeo classification with respect to the wavelength coverage and resolution. The near-infrared reflectance of 185 Eunike was exhibited in Barucci et al. (1994); nevertheless, the continuity between the visible and near-infrared region necessary for classification was lacking due to the absence of data in the 1–1.15 $\mu$m wavelength region. Therefore, the follow-up observations of four AcuA-spec asteroids are spectroscopically made, in this study. Additionally, observations on the 42 Isis, 128 Nemesis, and 476 Hedwig, which are members of the AcuA-spec, are made to check the reliability of the obtained reflectance.

Spectroscopic observations of the AcuA-spec asteroids were performed with a 1.88 m telescope at the Okayama Astrophysical Observatory, National Astronomical Observatory of Japan, National Institute of Natural Sciences in Okayama, Japan (MPC code 371; Coordinates: 133°35'38"E, 34°34'37"N; Altitude: 360 m) during February and August 2016. The asteroids were observed for four nights. The nightly observation details of the spectroscopy are summarized in table 1.

Spectroscopic data was obtained in the low dispersion mode of an infrared spectrograph with a low noise efficient detector (ISLE) instrument (Yanagisawa et al. 2006, 2008) attached to the f/18 Cassegrain focus of a 1.88 m telescope. The instrument detector had a 1024 × 1024 HAWAII HgCdTe array, which provided a 4.′3 × 4.′3 field-of-view, with a pixel scale of 0.″25. The ISLE spectroscopic system was composed of reflection gratings. Data was acquired by dividing the wavelength range into two wavelengths. The two spectral ranges determined with the YJH and HK filters were 0.95–

---

[2] Vernazza et al. (2016) was published, after the observations in this study.

Table 1. Observational circumstances of the AcuA-spec asteroids.

| Num | Name | Previous NIR observations[*] | Date (UT) [YY.MM.DD] | Start – End time [hr:min] | Airmass | Solar analogue |
|---|---|---|---|---|---|---|
| 94 | Aurora | | 2016.02.17 | 18:19 – 19:09 | 1.27 – 1.34 | SA104-335, HD111662 |
| | | | 2016.02.18 | 18:24 – 19:06 | 1.27 – 1.24 | HD115152 |
| 127 | Johanna | S10, V16 | 2016.02.17 | 17:16 – 17:38 | 1.08 – 1.10 | HD107146 |
| | | | 2016.02.18 | 15:05 – 15:29 | 1.15 – 1.20 | HD107146 |
| | | | | 19:32 – 19:50 | 1.25 – 1.42 | HD107146 |
| 185 | Eunike | B94, S10 | 2016.08.01 | 18:18 – 18:49 | 1.29 – 1.31 | SA115-271, HD224383 |
| 423 | Diotima | | 2016.02.17 | 11:36 – 15:23 | 1.10 – 1.30 | SA102-1081 |
| 42 | Isis | B88, C95, B02, D09 | 2016.04.19 | 14:32 – 14:53 | 1.32 – 1.37 | SA102-1081 |
| 128 | Nemesis | D09, S10 | 2016.04.19 | 15:03 – 15:24 | 1.31 – 1.32 | HD139287 |
| 476 | Hedwig | S10, H11 | 2016.04.19 | 12:03 – 13:13 | 1.28 – 1.44 | SA102-1081 |

[*]B88, B94, C95, B02, B04, D09, S10, H11, and V16 indicates Bell et al. (1988), Barucci et al. (1994), Clark et al. (1995), Burbine & Binzel (2002), Baudrand, Bec-Borsenberger, & Borsenberger (2004), DeMeo et al. (2009), Sykes et al. (2010), Howell et al. (2011) and Vernazza et al. (2016), respectively.

1.85 $\mu$m and 1.45–2.53 $\mu$m, respectively (figure 1). The H-band is a wavelength region, where the two spectral modes overlap. The ISLE in low resolution has a dispersion of approximately 6 Å per pixel in the Y and J band regions, and approximately 13 Å per pixel in the H and K band regions, in the wavelength direction. The length of the long slit was 4′ in the cross-wavelength direction. The orientation of the slit was fixed to the east–west direction. As the typical seeing size at the Okayama Astrophysical Observatory is about several seconds, data was obtained in the sidereal tracking mode and the slit direction was not perpendicular to the horizon; a 5″-wide slit was used in order to avoid slit loss, which leads to a production of incorrect spectra for asteroids and standards. The spectral resolutions in low dispersion, in the Y and J bands, and in the H and K bands were ∼120 Å and ∼260 Å respectively, using the 5″ slit,.

As standard stars, solar analogue stars listed in previous papers (e.g., Clark et al. 2009) or classified as G2V stars in the SIMBAD Astronomica database[3] were used. Observations between the asteroid and standard star were coordinated such that the airmass difference was less than 0.1, in each case. Data was obtained with an airmass lower than 1.4, to prevent airmass mismatch. The asteroids and standards were located in the center of the slit using a slit-viewer CCD. Spectroscopic data was obtained at two different locations referred to as the "A" and "B" positions on the detector to enable the subtraction of the atmospheric and thermal emissions. A single observation set was composed of four "A–B–A–B" spectra. As much as possible, the spectrum on the image was repeatedly placed on the same column. For signal acquisition within the linearity range of the detector, the total exposure time was split into individual exposures of 120 s. Dome flat-fielding frames in the YJH and HK bands were obtained each night. The wavelength calibration of the obtained frames was done with light from argon and xenon lamps, immediately after the spectral observations.

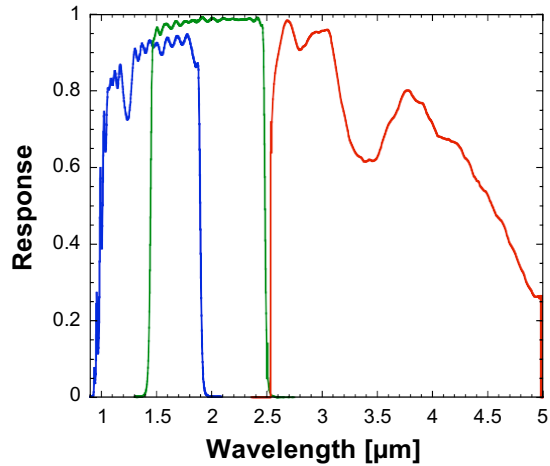

**Fig. 1.** Filter transmittance for the ISLE spectroscopic observations and the relative spectral response function (RSRF) of the spectroscopy mode for AKARI. The blue and green lines indicate the transparency of the YJH and HK ISLE filters, respectively. The red lines indicate the RSRF for the near-infrared spectroscopic observations, for AKARI. The RSRF data for AKARI was obtained from Shimonishi et al. (2010).

The typical seeing size on the four nights was ∼2". The weather condition during data acquisition changed from cloudy to photometric. Therefore, data under the photometric condition, as decided by the signal counts and sky values after the A-B subtraction, was only selected.

All data reduction was done using the image reduction and analysis facilities (IRAF) software. The spectra were reduced by subtracting the consecutive A and B images with a flat-fielding correction. The extraction of a 1-D spectrum from the 2-D images was performed using the *apall* task, with the IRAF software. The spectra were finally smoothed with the median filter technique, according to the wavelength resolution of the 5″ slit. The spectra of the solar analogues were treated using the same process as the asteroid spectra. The reflectance of the asteroids was obtained through spectrum of the asteroids to the spectrum of the solar analogue star. Individual spectra were finally averaged to obtain the final reflectance spectrum of each frame, which was normalized to 1.60 $\mu$m.

## 3 Spectroscopic results and classification of AcuA-spec asteroids

The AcuA-spec asteroid spectra obtained for the follow-up observations are presented in figure 2. The spectra for indicating the validity of the observations are also shown in figure 3. Where possible, visible data of large surveys such as the Zellner, Tholen, & Tedesco (1985), Xu et al. (1995), Bus & Binzel (2002a), and Lazzaro et al. (2004) were included. Near-infrared spectrophotometric and spectroscopic data acquired in the past are also depicted in the figures (Bell et al. 1988; Barucci et al. 1994; Clark et al. 1995; Burbine & Binzel 2002; DeMeo et al. 2009; Howell et al. 2011; Vernazza et al. 2016). In this study, the obtained spectra coincide with the previous spectra (see the asteroid 42 Isis, 128 Nemesis, and 476 Hedwig in figure 3, and 127 Johanna and 185 Eunike in parts of figure 2). This indicates the validity of the spectra in this study.

[3] http://simbad.u-strasbg.fr/simbad/

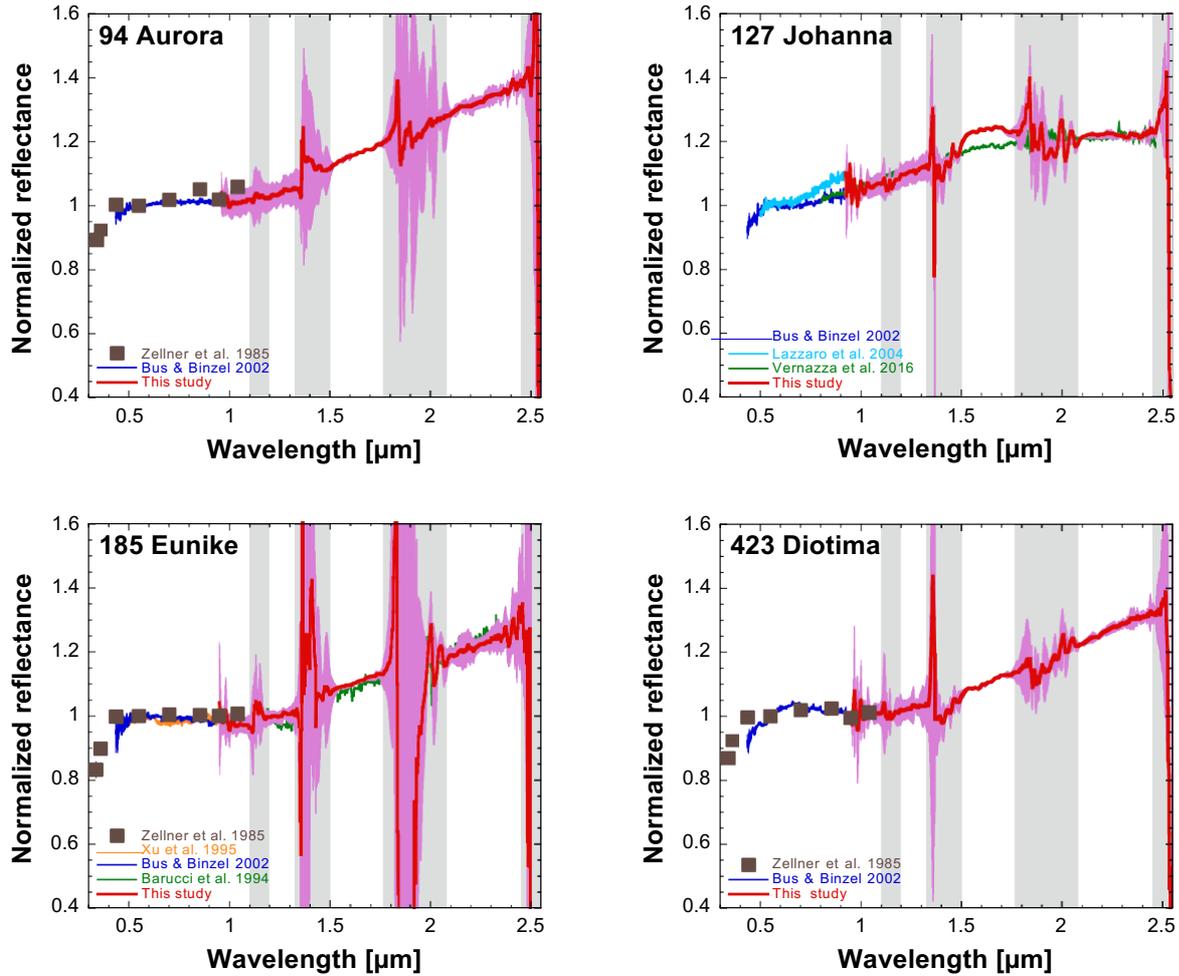

**Fig. 2.** Spectra for the AcuA-spec follow-up observations in the visible and near-infrared regions. The reflectance is normalized at 0.55 $\mu$m. The grey bars near 1.15, 1.4, 1.9, and 2.5 $\mu$m mark the strong absorption regions in the terrestrial atmosphere.

Of the 65 AcuA-spec asteroids, 39 asteroids were already classified, based on the Bus-DeMeo taxonomy (DeMeo et al. 2009). Subsequently, six asteroids were similarly classified with the taxonomy (Clark et al. 2009; Fornasier et al. 2010; Ockert-Bell et al. 2010; Neeley et al. 2014). As the Bus-Demeo taxonomic classification of 20 AcuA-spec asteroids, including the four asteroids in this study was not done previously, these asteroids are classified in this study. The data of 16 uncategorized asteroids were obtained from previous papers (see table 2). In order to classify the asteroids, based on the Bus-DeMeo taxonomy, a tool available on the world wide web[5] was used. The tool taxonomy (Binzel et al. in prep.)[6] was slightly updated from DeMeo et al. (2009). In this study, the obviously incorrect spectra in the wavelength region with a strong atmospheric absorption between the J and H bands, and between the H band and K bands were corrected by linear interpolation, using the correct spectra around them.

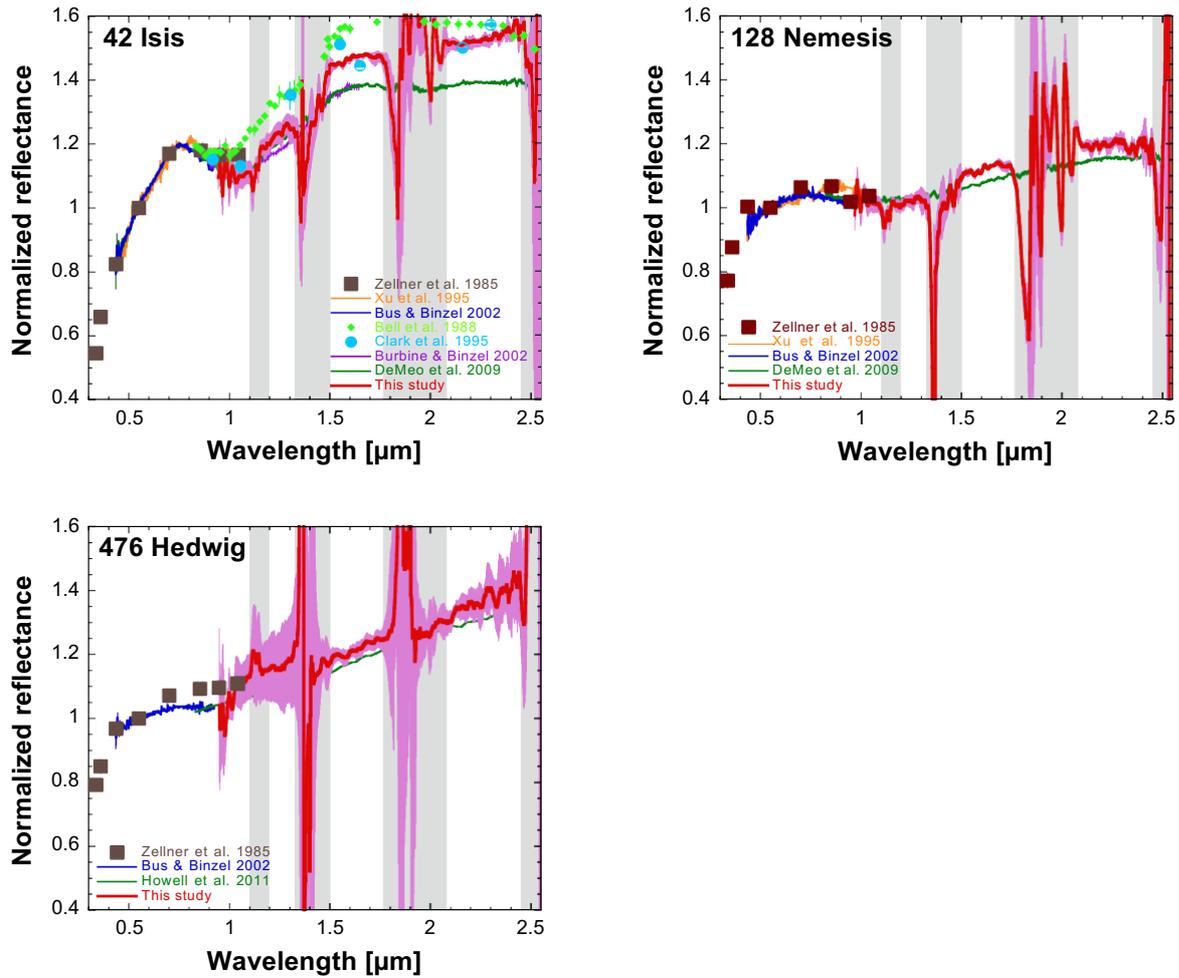

**Fig. 3.** Visible and near-infrared spectra of the AcuA-spec asteroids for the confirmation of the process in this study. The reflectance is normalized at 0.55 μm. The grey bars near 1.15, 1.4, 1.9, and 2.5 μm mark the strong absorption regions in the terrestrial atmosphere.

The classification results for all the AcuA-spec asteroids in the Bus-DeMeo taxonomy are shown in table 2. As planned, the C-, S-, X-complex, and end member asteroids were all obtained. However, it found that only the S-type asteroids in the S-complex were observed in the AcuA-spec asteroids. Sa- and Sv-type asteroids are small asteroids that could not be detected by spectroscopic observations with the AKARI; however, the Sq- and Sr-asteroids also could not be observed because the Sl-type (148 Gallia) and Sq type asteroids (33 Polyhymnia) in the Bus-Binzel taxonomy (Bus & Binzel 2002b) were altered in the S-type asteroid Bus-DeMeo classification. The Cg-, O-, and Q-type asteroids could not be observed because Q-type asteroids are mostly present in the near- Earth asteroids (see also figure 4) and two the 175 Andromache and 7088 Ishtar and only the 3628 Boznemcova are currently found as Cg- and O-type asteroids, respectively.

---

[4] http://smass.mit.edu/minus.html

[5] http://smass.mit.edu/busdemeoclass.html

[6] Binzel et al. (in prep.) added the Xn-type because they discovered a few asteroids that had a Nysa-like spectrum with a small feature at 0.9 μm. Xn-type has certain characteristics that are similar to Tholen E-type asteroids.

## 4  Combination of the Bus-DeMeo classification with the other physical characteristics

There is a possibility that the combination of the Bus-DeMeo taxonomy and the other physical characteristics may constrain the physical properties of the asteroids. The Bus-DeMeo classification for all the AcuA-spec asteroids was performed in the previous section; however, in addition to the classification data of the AcuA-spec asteroids, all the data in the accepted articles prior to this study, should be utilized to constrain each spectral type asteroid, using the Bus-DeMeo taxonomy from a statistical point of view. Therefore, the Bus-DeMeo classification, for use in this study, was extended to the accepted articles, prior to this study. The classification results, based on the Bus-DeMeo taxonomy, are quoted from DeMeo et al. (2009), Clark et al. (2009), Clark et al. (2010), Fornasier et al. (2010), Ockert-Bell et al. (2010), Fornasier, Clark, & Dotto (2011), Ostrowski et al. (2011), de León et al. (2012), Gietzen et al. (2012), DeMeo, Binzel, & Lockhart (2014), Ieva et al. (2014), Neeley et al. (2014), Polishook et al. (2014), Perna et al. (2016), Lucas et al. (2017), and the previous section of this study. 767 asteroids were already classified using the Bus-DeMeo taxaonomy.

The Bus-DeMeo taxonomy uses asteroid reflectance spanning from 0.45–2.5 $\mu$m to classify asteroids. Combined with the other observational information such as the geometric albedo, polarization, radar, and UBV, it was attempted to extract the physical properties of the classified asteroids in the Bus-DeMeo taxonomy.

Before the combination, the classifications of 23 asteroids, including 17 asteroids for polarimetric observations, five high albedo asteroids, and a special asteroid were included. In order to correlate the polarimetric parameter with the Bus-DeMeo taxonomy, the classification of asteroids, based on the polarization and spectral observations, with a range from the visible to near-infrared wavelength region was done. As the Xn-type was newly included in Binzel et al. (in prep.), the classification of eight Tholen E-type asteroids in Clark et al. (2004a) was performed. The spectral types of five high albedo asteroids were included. Certain spectrophotometer data, below the wavelength resolution, necessary for executing the Bus-DeMeo classification were interpolated using linear interpolation. As the data from Clark et al. (1995) does not contain values more than 2.30 $\mu$m, taxonomic classification was performed by assuming that the reflectance over 2.35 $\mu$m was the same as that at 2.30 $\mu$m. The taxonomic results for 23 asteroids are presented in table 3. A total of 790 asteroids were classified, based on the Bus-DeMeo taxonomy. The number of the asteroids are approximately an order greater than that of the AcuA-spec asteroids.

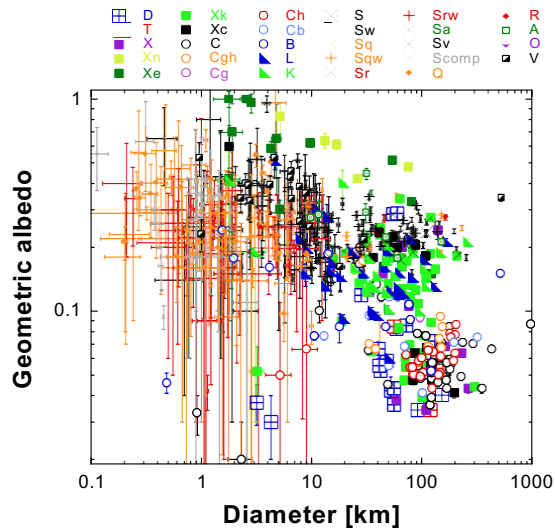

**Fig. 4.** Relationship between the size and geometric albedo of the asteroids classified by Bus-DeMeo taxonomy.

Among the classified asteroids, the geometric albedo values of 617 asteroids are available in literature (Bell, Hawke, & Brown 1988; Campins et al. 2009; Cruikshank et al. 1991; Delbó et al. 2003, 2011; Emery et al. 2014; Kraemer et al. 2005; Mainzer et al. 2011b, 2012, 2014; Masiero et al. 2011, 2012, 2014; Morrison 1977; Mueller et al. 2011; Müller et al. 2004, 2014; Müller, Hasegawa, & Usui 2014; Nugent et al. 2015, 2015; Tedesco et al. 2002; Thomas et al. 2011; Trilling et al. 2010, 2016; Usui et al. 2011; Veeder et al. 1989; Wolters et al. 2008; the wolrd wide web page of NEOSurvey[7]). The detection limit of the classified main-belt asteroids is ∼50 km (figure 4). The diameter of the smallest AcuA-spec asteroid is also 50 km. The almost classed near-Earth asteroids belong to the S-complex and Q-type asteroids. The albedo distribution of the classified asteroids in the Bus-DeMeo taxonomy is shown in figure 5. The geometric albedo data for the classified asteroids is approximately greater than that of the AcuA-spec asteroids by a factor of eight.

The UBV photometric data of 289 classified asteroids are quoted from Barucci, Di Martino, &

---

[7] http://nearearthobjects.nau.edu/neosurvey/

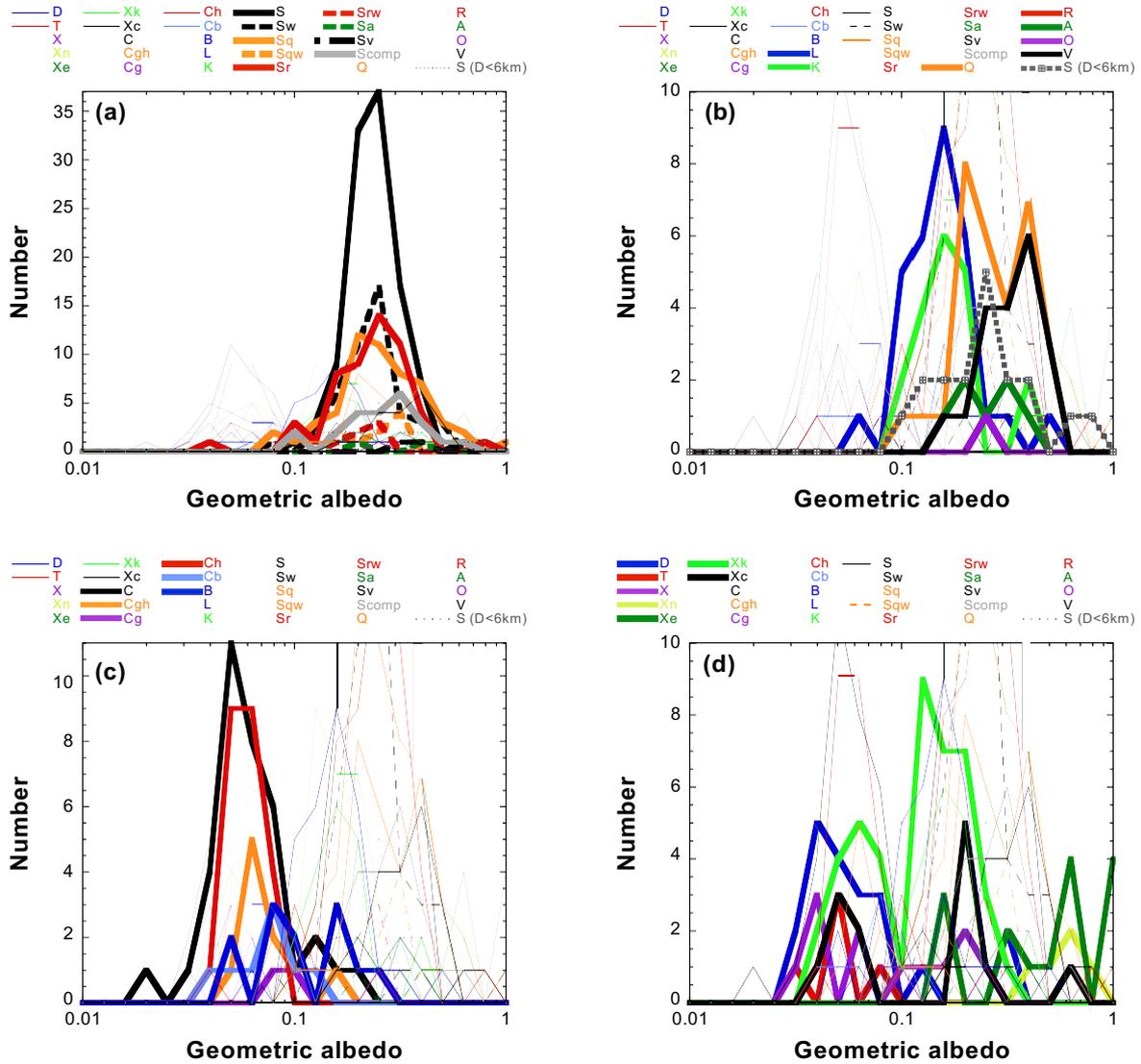

**Fig. 5.** Histogram of the geometric albedos of asteroids, based on the Bus-DeMeo taxonomy. The distribution for S-complex, end-member, C-complex, and D- and D-complex asteroids are identifiably shown as highlights in (a), (b), (c), and (d), respectively.

Fulchignoni (1994), Tedesco (2005), Thomas-Osip et al. (2008), and Toth (1997). The distributions of the U-B data for each taxonomic asteroid are depicted in figure 6. The UBV data for the classified asteroids is greater than that of the AcuA-spec asteroids by a factor of four approximately.

The polarimetric observation results of 86 asteroids using the Bus-DeMeo taxonomy were referred from (Belskaya et al. 2017; Cellino et al. 2015; Cellino et al. 2016; Gil-Hutton, Cellino, & Bendjoya 2014); the radar values for 74 classified asteroids were obtained from Neese, Benner, & Ostro (2012). The polarimetric and radar values of the classified asteroids are shown in figures 7 and 8, respectively. 28 polarimetric and 26 radar data are available for the AcuA-spec asteroids. The polarization and radar data for the classified asteroids are greater than those for the AcuA-spec asteroids by a factor of three, approximately.

## 5 Discussion

### 5.1 Nature of the asteroids using the Bus-DeMeo taxonomy

Focusing on the albedo distribution of the S-complex asteroids, there are certain differences in the peak positions, but the distributions almost coincide (figure 5). It is difficult to specify the S-, Sq-, or Sr-types within the S-complex, from the geometric albedo. The albedo distribution of Q-type asteroids is slightly higher than that of the S-complex asteroids (figure 5). However, Pravec et al. (2012) pointed out that the increase in albedos with decreasing asteroid size for diameters less than ∼30 km as shown in Mainzer et al. (2011a) can be attributed to the systematic bias arising from the use of absolute magnitudes. As the Q-type asteroids in this study are distributed in this range, there is a possibility that this bias influences the observations. However, the albedo distribution of S-type asteroids in the range up to 6 km, where the Q-type asteroids are distributed, is slightly lower than that of the Q-type asteroids (figure 5). Binzel et al. (2004) insisted that the Q-type and S-complex of the near-Earth asteroids were from the effectively identical main-belt source region, as would be expected, if they had related origins, using massive spectroscopic data. From the analysis of samples brought by the Hayabusa spacecraft from the asteroid, 25143 Itokawa, classed as an Sqw-type asteroid (see table 3), it is determined that the composition is identical to those of LL chondrite (Nakamura et al. 2011; Nakamura et al. 2014). This demonstrates that the surfaces of S-complex asteroids are altered by the space weathering process from that of the ordinary chondrites, whose spectrum is similar to that of Q-type asteroids. From the laboratory experimental space weathering results, regardless of the cosmic ray origin and interplanetary dust collision formation, the albedo value of the sample reduces in comparison to that before exposure (Sasaki et al. 2001; Brunetto & Strazzulla 2005). Thus, the high distribution of the albedo values of the Q-type asteroids, compared to those of the S-type asteroids, is consistent with previous studies.

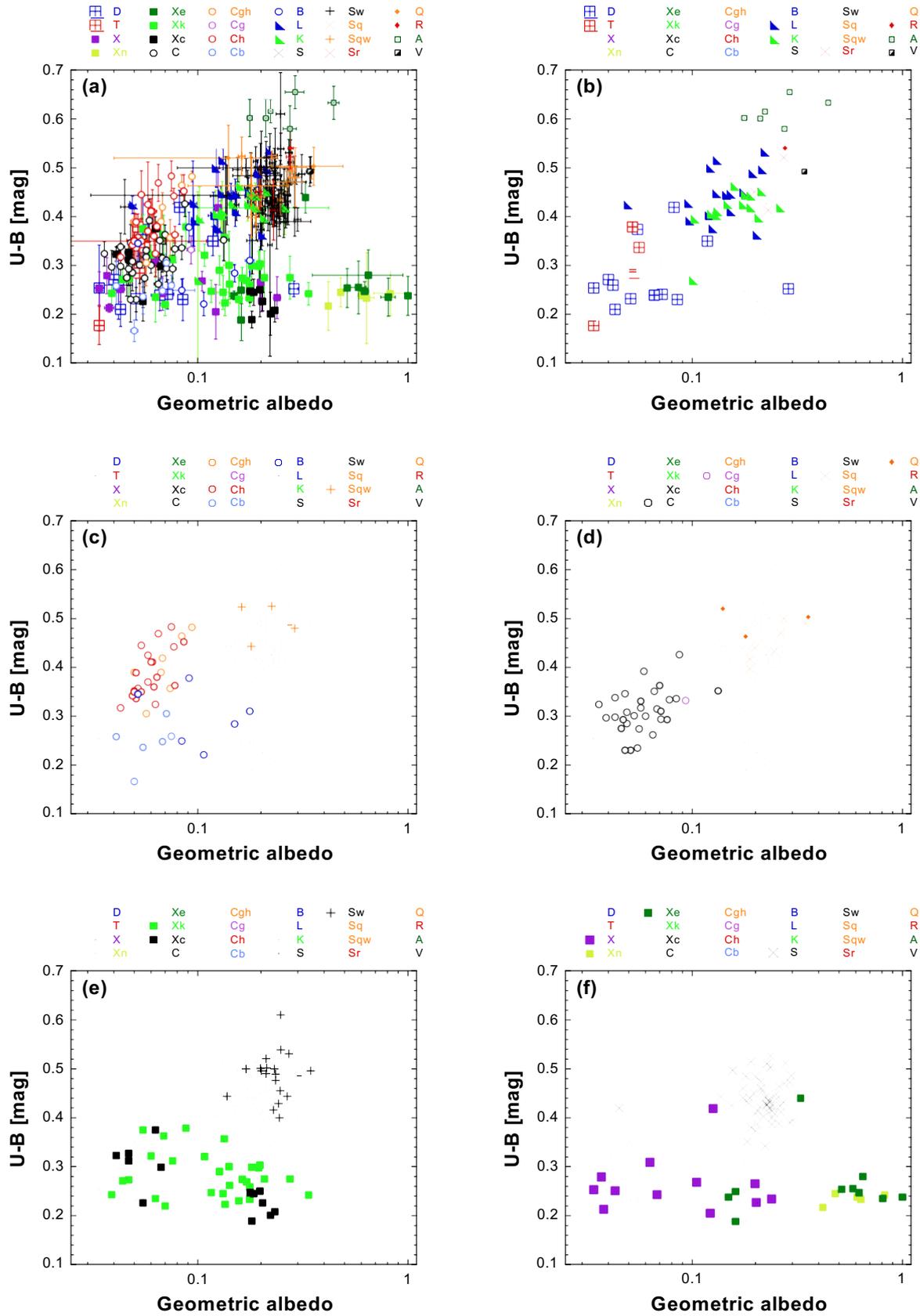

**Fig. 6.** Relationship between the geometric albedo and the U-B colors of the classified asteroids based on the Bus-DeMeo taxonomy. (a) all spectral type asteroids. (b) D-, T-, L-, K-, Sr-, R-, A-, and V-type asteroids. (c) Cgh-, Ch-, Cb-, B-, and Sqw-type asteroids. (d) C-, Cg-, Sq-, and Q-type asteroids. (e) Xk-, Xc-, and Sw-type asteroids. (f) X-, Xn-, Xe-, and S-type asteroids.

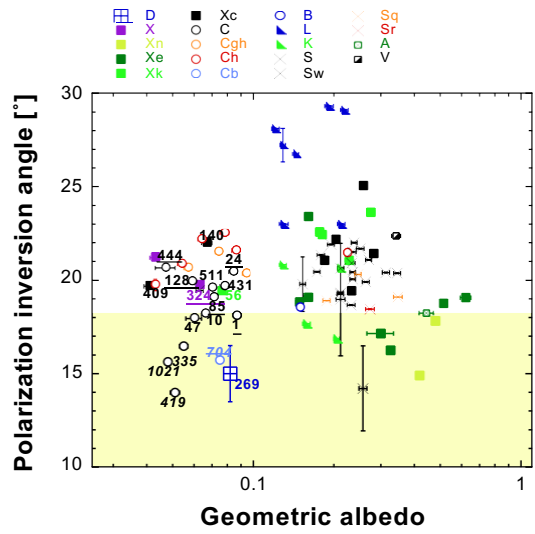

**Fig. 7.** Distribution of the geometric albedo, U-V color, and inversion angle from the polarimetric observation of the asteroids, in the Bus-DeMeo taxonomy. The numbers in the figure indicate the asteroid number. Underlined numbers indicate that there is a radar observation value, whereas italicized numbers indicate Tholen F-type asteroids.

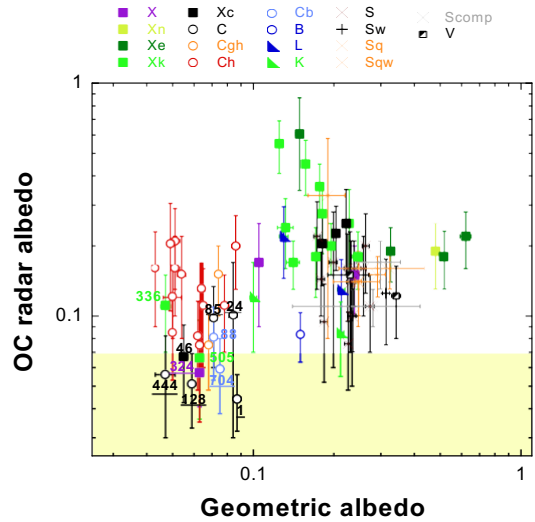

**Fig. 8.** Comparison between the asteroidal geometric albedos and opposite circular (OC) radar albedos with the Bus-DeMeo taxonomy. Underlined numbers indicate that there is a polarization observation value for the asteroid.

The distribution of the albedo values of the K- and L-type asteroids is lower than that of the S-complex asteroids and higher than that of the C-complex asteroids (figure 5). The U-B distributions show similar trends (figure 6). It is suggested that the K-type asteroids are the source of the CK, CO, and CV chondrites (Clark et al. 2009); certain L-type asteroids are the parent bodies of the CV3 chondrites (Sunshine et al. 2008). Cellino et al. (2006) and Gil-Hutton et al. (2008) determined that L-type asteroids have an unusual polarimetric behavior characterized by huge inversion phase angles. Figure 7 also shows that some the L-type asteroids have very high polarimetric inversion angles (~30°). However, for some of the L-type asteroids, the inversion angle is not extremely large. Sunshine et al. (2008) showed that the Calcium- and aluminum-rich inclusions (CAIs) in the CV3 chondrites have strong absorption features at 2.0 $\mu$m, attributed to the iron oxide-bearing aluminous spinel. The spinel commonly incorporates small, usually present in inclusions within a dark matrix in these meteorites. Gil-Hutton et al. (2008) propose that the L-type asteroid surface may be a coarse regolith with a dark matrix, mixed with smaller white inclusions such as CAIs and that this mixture of two components with different albedos can explain the large inversion angle seen in the L-type asteroids. The geometric albedo values of the CV3 and CK4 chondrites (Hiroi et al. 1994) are within the range of those of the K- and L-type asteroids. This implies that the surface compositions of L-type asteroids with large inversion angles, and those of the K- and L-type asteroids with normal inversion angles are compatible with those of the CV3 and CK, CO chondrites, respectively.

It is traditionally known that the albedo value of X-complex asteroids diversifies. In the Tholen taxonomy, the X-type is classified as E-, M-, and P-type asteroids, using the albedo values (Tholen 1994). However, even with the Bus-DeMeo taxonomy, the geometric albedo distribution of X-complex asteroids, except for the Xn-type asteroid, does not exist collectively (figure 5). The X- and Xk-type asteroids have bimodal distributions between a low (less than 0.1) and medium albedo (0.1–0.3). The albedo distribution of Xe-type asteroids is from medium to high (more than 0.3). The histogram peak of the X-type asteroids is trimodal.

The distributions of Xk- and Xe-type asteroids with medium albedos is close to those of the K- and L-asteroids. Thus, Xk- and Xe-type asteroids with intermediate albedos may be analogous to CK, CO, and CV chondrites, considering the K- and L-type asteroids. However, Vernazza et al. (2009) argued that Xk-type asteroids with medium albedos are the parent bodies of mesosiderites. Xc-type asteroids with intermediate albedos have physical properties compatible with those of enstatite chondrite and aubrites (Vernazza et al. 2009; Vernazza et al. 2011). Shepard et al. (2010), Ostrowski et al. (2011), and Neeley et al. (2014) suggested that X-complex asteroids with medium albedos are linked to stony-iron meteorites and high-iron carbonaceous chondrites, CV and CR chondrites, and iron meteorite and enstatite chondrites. The U-B colors of X-complex asteroids are not correlated with their geometric albedos (figure 6). From the radar albedo results, which are beneficial for estimating the range surface density for the penetration depth of the radar wavelength (e.g., Ostro et al. 2002), some Xk- and Xe-type asteroids with high radar albedos, analogous to high-density surface layers such as iron and stony-iron meteorites are present (figure 8). However, the values of the radar albedo of the X-complex asteroids, excluding the high-density asteroids, are almost constant. It is difficult to identify analogues of X-complex asteroids with medium albedos, except for the high-radar-albedo asteroids, using the reflectance, geometric albedo, U-B color, polarization, and radar albedo (figures 5 – 8).

Although the albedos of the C-, Ch-, and Cgh-type asteroids are distributed almost in the same location, the Cb- and B-type asteroids have slightly different albedo distributions compared to those of the C-, Ch-, and Cgh-type asteroids (figure 5). Moreover, the albedo distributions of X-complex asteroids with lower albedos, and those of the D- and T-type asteroids coincide with those of C-type asteroids (figure 5). The U-B distributions of the Ch- and Cgh-type asteroids are at a different location than those of the C- and Cb-type asteroids, the D- and T-type (D-complex), and the X-complex asteroids with low geometric albedos (figure 6).

It is to be noted that the spectra of the Cb- and B-type are similar to those of the CM2, the unusual CM2 chondrites, the CV3, and the CK4 chondrites, respectively (Clark et al. 2010; de León et al. 2012). Typical geometric albedos of meteorites and interplanetary dust particles (IDPs) are less than 0.1 (Bradley et al. 1996), whereas those of the CV3 and CK4 chondrites range between 0.1–0.2 (Hiroi et al. 1994). From a comparison of the asteroid spectra and the IDPs between the visible and mid-infrared wavelength region, Vernazza et al. (2015, 2017) specified that the C-, Cb-, Cg-, and B-type asteroids, and the D- and X-complex asteroids with lower albedos are related to anhydrous IDPs. Meanwhile, the Ch- and Cgh-type asteroids are considered to have the same composition as that of CM chondrites (Vernazza et al. 2015, 2016). Main-belt comets, which is expected to have anhydrous minerals from the existence of volatile ices, have spectra of C- and B-type (Licandro et al. 2011; Rousselot, Dumas, & Merlin 2011; Licandro et al. 2013; Fernández et al. 2017; Snodgrass, Yang, & Fitzsimmons 2017). Bradley et al. (1996) and Noguchi et al. (2017) discussed that the dark X- and D-complex asteroids may be the more plausible parent bodies of IDPs, through an analysis of the IDPs obtained from the stratosphere and the Antarctic snow, respectively. These show that the C-, Cb-, B-type, and the X- and D-complex asteroids with albedos less than 0.1, are related to the composition of anhydrous IDPs.

With geometric albedos more than 0.1, the value of the OC radar albedos for S-complex and X-complex asteroids without metallic surface candidates are distributed in the range, 0.08–0.3 (figure 8). Although the radar albedos of the Ch- and Cgh-type asteroids fall within this range, some of the C-, Cb-type, and X-complex asteroids with low geometric albedos, have radar albedos below this range. In case of asteroids with OC radar albedos in the range of 0.08–0.3, it is expected that they contain regoliths of chondritic or achondritic material (Shepard et al. 2010). On the other hand, the OC radar albedo values for most comets are less than 0.07 and the results of the comet radar albedos indicate that the comet nuclei have a regolith surface with a mixture of chondritic material and dry snow (Harmon et al. 2004). These suggest that the surface of asteroids with OC radar albedos less than ∼ 0.07 are also have compositions comparable to those of comets. Thus, the surface of these asteroids is expected to be made up of anhydrous chondritic materials and water ice that are not advanced in the aqueous alteration.

Based on the polarimetric inversion angle of low albedo asteroids, they are divided into two groups: one asteroid group, whose angle is approximately more than 20°, composed of Ch-, Cgh-, and certain C-type and X-complex asteroids, and another asteroid group with an angle less than 19°, composed of C-, Cb-, and D-type asteroids (figure 7). It is known that some of the Tholen F-type asteroids have low polarimetric inversion angles (Belskaya et al. 2005); the F-type asteroids correspond to them. However, not all of them are Tholen F-type asteroids (see figure 7). Belskaya et al. (2005) proposes two feasible mechanisms for low polarimetric inversion angles. One is that the surface optical homogeneity is caused owing to the carbon deposits, which are metamorphized from the organic matter at the top surface layer. The other is that it is caused by dark small particles for adding small amounts of bright particles. Steigmann (1993) demonstrate that the polarimetric inversion angle become less than 18°, under a frost layer, with a large quantity of dark particles. The polarimetric characteristics of 2P/Encke and 133P/Elst-Pizarro have the same trend as that of Tholen F-type asteroids (Bagnulo et al. 2010). This implies the presence of water ice on the surface of dark asteroids with low inversion angles.

Using ground-based observational data in the 2–4 $\mu$m range, Takir & Emery (2012) classified dark asteroids into several types. Most observed Ch- and Cgh-type type asteroids belong to asteroids, whose shape-type is attributed to hydrated minerals (Rivkin et al. 2015); rounded-type asteroids, attributed to water ice, are composed of C-type, and X- and D-complex asteroids (Takir & Emery 2012). The 3-$\mu$m band absorption feature of water ice is highly sensitive and a thickness of more than 0.1 $\mu$m results in band saturation (Rivkin et al. 2015). Therefore, even a small amount of water ice can be detected in this band.

3-$\mu$m band absorption has been attributed not to water ice but to ammonium-bearing phyllosilicates (King et al. 1992), brucite (Beck et al. 2011), and a molecular mixture, where water molecules are uniformly mixed in the regolith particle (Clark 2009). These judgments are difficult with the 3-

$\mu$m absorption band only. However, the possibility of the existence of water ice on the asteroids is suggested, based on this study. For example, it is known that 24 Themis has water ice on the surface (Campins et al. 2010; Rivkin & Emery 2010). However, the 24 Themis, similar to the less aqueously altered meteorites, based on the emissivity in the mid-infrared wavelength region (Landsman et al. 2016), does not have a low radar albedo and inversion angle. There may be no-water ice particles such as ammonium-bearing phyllosilicates and very few water ice particles in the surface layer of the 24 Themis. 704 Interamnia is a shape-type asteroid (Takir & Emery 2012) but has a surface variation between the shape- and rounded-types (Rivkin et al. 2016); It has a low radar albedo and inversion angle. These facts suggest that water ice exists, to a certain extent, owing to the depth, which the radar wave can reach herein. Thus, the approach with respect to the various physical information, based on the Bus-DeMeo taxonomy, is beneficial for identifying the constituent substances on asteroid surfaces.

5.2 Application to AcuA-spec asteroids

As a result of including the information on the geometric albedo, U-band photometry, radar albedo, and polarimetry to the Bus-DeMeo taxonomy, the following were concluded in the previous chapter.

(1) S-complex and Q-type asteroids are related in terms of the surface compositions.
(2) L-type asteroids with large inversion angles have CV3 chondrite surfaces.
(3) The composition of Xk- and Xe-type asteroids, having high radar albedos, is compatible with those of iron- and stony-iron meteorites.
(4) Dark C-complex without the Ch- and Cgh-type, X-complex, and D-complex asteroids, are composed of anhydrous chondritic materials and water ice that are not advanced in the aqueous alteration.

With respect to the above, we consider the results from the AcuA-spec asteroids. The spectroscopic results of the AcuA-spec asteroids are expected to identify the absorption bands of the hydrated minerals (~2.7 $\mu$m), which are opaque owing to the atmosphere and the water in the minerals ~2.95, $\mu$m) positioned at the edge of the atmospheric transmission (Rivkin et al. 2015). Therefore, we expect elucidation on the origin and evolution of the hydrated minerals of the AcuA-spec asteroids, through the Bus-DeMeo taxonomy.

Q-type asteroids are not observed in the AcuA-spec, but there is a possibility of constraining the space weathering for S-complex asteroids. It is considered that hydrated minerals originate from aqueous alteration or implantation of solar wind protons (Rivkin et al. 2015). It may be possible to specify the source of hydrated mineral formation by the 3-$\mu$m observations of the AcuA-spec asteroids. Rivkin et al. (2018) argue that hydrated features are detected on the surface of the Sw-type asteroid, 433 Eros and the S-type asteroid, 1036 Ganymed, and suggest that the implantation of the solar wind protons could have created sufficient asteroid hydrated minerals. On the other hand, it has not been confirmed, if hydrated mineral formation is possible by the solar wind in the main-belt, owing to which weak absorption is expected. Absorption confirmation of the hydrated minerals on the surface layer of S-complex asteroids in the AcuA-spec asteroids is expected. Among the AcuA-spec asteroids, nine S-type and two Sw-type asteroids were detected. If confirmed, a continuum slope from 0.45–2.5 $\mu$m influenced by the space weathering effect may be beneficial for cause constraint.

Among the AcuA-spec asteroids, the 42 Isis, 387 Aquitania, and 9 Metis were acquired as K-type, L-type with a large inversion angle, and L-type without a large inversion angle, respectively. The 3-$\mu$m observations of these asteroids have not yet been done. The presence or absence of hydrated minerals may be beneficial for constraining the sources of these asteroids.

Landsman et al. (2015) confirmed that the 216 Kleopatra, which is considered to be composed of iron-rich material, has an absorption of 3-$\mu$m and proposed that it was brought about by the collision of a hydrated mineral-rich object on its surface. The exogenic contamination proposal for 3-$\mu$m absorption on asteroidal surface was presented by Hasegawa et al. (2003) and confirmed by the surface of the differentiated asteroid, 4 Vesta, by the Dawn spacecraft (McCord et al. 2012). On the other hand, Hardersen et al. (2005) suggests that the surface of the 216 Kleopatra is attributed to iron-poor orthopyroxenes and iron material. This does not deny that the 3-$\mu$m absorption may be caused by a solar wind origin. Hydrated minerals on the lunar surface were detected by the Cassini (Clark 2009), Deep Impact (Sunshine et al. 2009), and the Chandrayaan-1 spacecraft (Pieters et al. 2009). The minerals, which are formed by solar wind implantation appear at the position absorption features near 2.8–3.0 $\mu$m. Similar absorption was also detected in the soil of Apollo 16 and 17 (Ichimura et al. 2012). On the other hand, phyllosilicate absorption caused by aqueous alternation in meteorites, appear in the vicinity of 2.7 $\mu$m (e.g., Rivkin et al. 2015). The absorption position may be slightly different, depending on the absorption origin of the hydrated mineral. Among the AcuA-spec asteroids, the 2 Xk- and 1 Xe-type asteroids with high radar albedos are the 16 Psyche, the target of the NASA Discovery Program, Psyche mission (Elkins-Tanton et al. 2017), the 69 Hesperia, and the 216 Kleopatra. It may be possible to constrain the origin of the 3-$\mu$m absorption because AcuA-spec asteroids can be seen directly in the range, 2.5–2.85 $\mu$m.

The strength of the absorption feature of the hydrated minerals in asteroids composed of mainly chondritic IDPs is significant for determining the correlation with the degree of aqueous alternation. In the AcuA-spec asteroids, 11 C-type, 1 Cb-type, 1 B-type, 3 Ch-type, 2 Cgh-type, 3 X-type, 2 Xc-type, 5 Xk-type, 1 T-type, and 2 D-type dark asteroids were observed. In particular, the 1 Ceres, 46 Hestia, 128 Nemesis, and the 704 Interamnia were identified as dark asteroids with low radar albe-dos. The 1 Ceres, 10 Hygiea, 419 Aurelia, and 704 Interamnia are established as dark asteroids with low inversion angles. By combining the spectroscopic results of the AcuA-specs and the other clues in the radar albedos, polarimetric inversion angles, and the emissivity in the mid-infrared wavelength region, further constraints on the asteroid compositions are expected.

Most meteorites, except meteorite falls, are considered to have been weathered by the Earth because considerable time has passed, since they fell on the Earth. Most of the meteorites have the absorption features of a wide range of adsorbed water in the vicinity of 2.95 $\mu$m. Various studies have been conducted to eliminate the weathering effect (Beck et al. 2010; Takir et al. 2012). However, originally, some adsorbed water may have existed and it is difficult to determine the absorption of water, using only meteorite data. AcuA-spec data can assist in proving the validity of these studies. If there is an absorption of internal water at 2.95 $\mu$m in conjunction with water ice at 3.1 $\mu$m, it may restrict the surface evolution of asteroids. Although internal water and absorption water may be derived from the interior of the asteroid, there are possibilities of being exogenic as well, as in the case of the hydrated minerals. These restrictions may be expected, based on the presence or absence and the depth of the absorption bands.

## 6 Summary

The spectroscopic observations of AcuA-spec asteroids extending between 1–2.5 $\mu$m not acquired by near-infrared prior to this study, were carried out. All the AcuA-spec asteroids, including those from which the spectra were obtained previously, as well as the asteroids obtained this study were classified based on the Bus-DeMeo taxonomy. The classification is beneficial for considering the 3-$\mu$m results of the AcuA-spec asteroids.

In addition, the analogue estimation of extraterrestrial material for each asteroid type was performed by combining the classification results with other observational data. Based on the results of the polarization, radar, and the 3-$\mu$m observations on the Bus-DeMeo taxonomy, the compositions of some of the asteroids with a C-complex, except the Ch- and Cgh-types, the dark X-complex, and the D-complex, are linked to that of chondritic IPDs with water ice. This result supports the proposal of Vernazza et al. (2015, 2017).

A combination of the Bus-DeMeo classification for AcuA-spec asteroids with the estimation using other physical information such as the radar albedo, the inversion angle from polarimetry, and the emissivity in the mid-infrared wavelength region will be beneficial for material information constraints, from the AcuA-spec asteroid observations.

## Acknowledgments

We are grateful to Dr. Francesca E. DeMeo, Dr. Ellen S. Howell, Dr. Driss Takir, and Dr. Pierre Vernazza for sharing the valuable near-infrared spectra of the AcuA-spec asteroids. Part of the data utilized in this publication was obtained and made available by the MIT-UH-IRTF Joint Campaign for NEO Reconnaissance. The IRTF is operated by the University of Hawaii under the cooperative agreement no. NCC 5-538, with the National Aeronautics and Space Administration, Office of Space Science, Planetary Astronomy Program. The MIT component of this work is supported by the NASA grant 09-NEOO009-0001 and by the National Science Foundation, under grant nos. 0506716 and 0907766. The taxonomic type results presented in this work were determined using a Bus-DeMeo taxonomy classification web tool by Dr. Stephen M. Slivan, developed at MIT with the support of the National Science Foundation Grant 0506716 and NASA Grant NAG5-12355. This study has utilized the SIMBAD database, operated at CDS, Strasbourg, France, and the JPL HORIZON ephemeris generator system, operated at JPL, Pasadena, USA. We are grateful to Professor Keiji Ohtsuki and Ms. Mako Imou for their support. We would like to express our gratitude to the staff members of the Okayama Astrophysical Observatory for their assistance. We thank an anonymous reviewer for their careful and constructive reviews, which helped us improve the manuscript significantly. DK is supported by the Optical & Near-Infrared Astronomy Inter-University Cooperation Program, the MEXT of Japan. This study was supported by the JSPS KAKENHI Grant Numbers JP15K05277 and JP17K05636, and by the Hypervelocity Impact Facility (former facility name: the Space Plasma Laboratory), ISAS, JAXA.

**Table 2.** Bus-DeMeo taxonomy for AcuA-spec asteroids

| Num | Name | B-D tax | Ref tax* | Ref spec* |
|---|---|---|---|---|
| 1 | Ceres | C | D09 | |
| 2 | Pallas | B | D09 | |
| 4 | Vesta | V | D09 | |
| 5 | Astraea | S | D09 | |
| 6 | Hebe | S | This study | V14 |
| 7 | Iris | S | D09 | |
| 8 | Flora | Sw | D09 | |
| 9 | Metis | L | This study | Z85 + L04 + R16 |
| 10 | Hygiea | C | D09 | |
| 13 | Egeria | Ch | D09 | |
| 16 | Psyche | Xk | D09 | |
| 21 | Lutetia | Xc | D09 | |
| 22 | Kalliope | X | D09 | |
| 24 | Themis | C | D09 | |
| 33 | Polyhymnia | S | D09 | |
| 40 | Harmonia | S | D09 | |
| 42 | Isis | K | D09 | |
| 44 | Nysa | Xn | This study | SMASS |
| 46 | Hestia | Xc | This study | C04b |
| 49 | Pales | Ch | D09 | |
| 50 | Virginia | Ch | D09 | |
| 51 | Nemausa | Cgh | D09 | |
| 52 | Europa | C | D09 | |
| 55 | Pandora | Xk | D09 | |
| 56 | Melete | Xk | D09 | |
| 64 | Angelina | Xe | D09 | |
| 65 | Cybele | Xk | D09 | |
| 69 | Hesperia | Xk | D09 | |
| 79 | Eurynome | S | This study | SMASS |
| 81 | Terpsichore | C | This study | SMASS |

**Table 2.** (Continued)

| Num | Name | B-D tax | Ref tax* | Ref spec* |
|---|---|---|---|---|
| 87 | Sylvia | X | D09 | |
| 89 | Julia | Sw | This study | V14 |
| 92 | Undina | Xk | D09 | |
| 94 | Aurora | C | This study | This study |
| 106 | Dione | Cgh | D09 | |
| 121 | Hermione | Ch | This study | SMASS |
| 127 | Johanna | Ch | This study | This study |
| 128 | Nemesis | C | D09 | |
| 129 | Antigone | Xk | O10 | |
| 135 | Hertha | Xk | O10 | |
| 140 | Siwa | Xc | This study | T12 |
| 145 | Adeona | Ch | This study | V16 |
| 148 | Gallia | S | This study | V14 |
| 153 | Hilda | X | D09 | |
| 161 | Athor | Xc | F10 | |
| 173 | Ino | Xk | C09 | |
| 185 | Eunike | C | This study | This study |
| 216 | Kleopatra | Xe | D09 | |
| 246 | Asporina | A | D09 | |
| 250 | Bettina | Xk | D09 | |
| 308 | Polyxo | T | D09 | |
| 336 | Lacadiera | Xk | N14 | |
| 349 | Dembowska | R | D09 | |
| 354 | Eleonora | A | D09 | |
| 361 | Bononia | D | This study | T12 |
| 371 | Bohemia | S | D09 | |
| 387 | Aquitania | L | D09 | |
| 419 | Aurelia | C | C09 | |
| 423 | Diotima | C | This study | This study |
| 451 | Patientia | C | This study | T12 |
| 476 | Hedwig | Xk | This study | B02 + H11 |

**Table 2.** (Continued)

| Num | Name | B-D tax | Ref tax[*] | Ref spec[*] |
|---|---|---|---|---|
| 511 | Davida | C | This study | T12 |
| 532 | Herculina | S | D09 | |
| 704 | Interamnia | Cb | This study | T12 |
| 773 | Irmintraud | T | D09 | |

[*]References for taxonomy and spectrum: Z85, Zellner, Tholen, & Tedesco (1985); B02, Bus & Binzel (2002a); C04b, Clark et al. (2004b); L04, Lazzaro et al. (2004); C09, Clark et al. (2009); D09, DeMeo et al. (2009); F10, Fornasier et al. (2010); O10, Ockert-Bell et al. (2010); H11, Howell et al. (2011); T12, Takir & Emery (2012); V14, Vernazza et al. (2014); R16, Reddy & Sanchez (2016); V16, Vernazza et al. (2016); SMASS, Database of The MIT-UH-IRTF Joint Campaign for NEO Spectral Reconnaissance[4].

**Table 3.** Bus-DeMeo taxonomy for polarimetric and special asteroids

| Num | Name | B-D tax | Ref spec[*] | Previous B-D tax[*] |
|---|---|---|---|---|
| 12 | Victoria | L | B02 + R16 | |
| 47 | Aglaja | C | Y10 | |
| 71 | Niobe | Xe | B02 + C04b | |
| 113 | Amalthea | Sr | SMASS | |
| 124 | Alkeste | S | V14 | |
| 138 | Tolosa | Sw | Ha06 | |
| 172 | Baucis | L | B02 + C95 | |
| 197 | Arete | Sw | B02 + C95 | |
| 230 | Athamantis | S | V14 | |
| 324 | Bamberga | X | T12 | |
| 376 | Geometria | Sw | B02 + B88 | |
| 377 | Campania | Cgh | V16 | |
| 409 | Aspasia | Xc | B02 + C04b | |
| 584 | Semiramis | Sw | B02 + B88 | |
| 796 | Sarita | Xk | Ha11 | |
| 980 | Anacostia | L | S08 | |
| 2867 | Steins | Xe | B05 | |
| 214 | Aschera | Xn | C04a | Cgh (D09) |
| 317 | Roxane | Xn | C04a | |
| 1103 | Sequoia | Xn | C04a | Xk (C09) |
| 1251 | Hedera | Xn | C04a | |
| 2048 | Dwornik | Xe | C04a | |
| 25143 | Itokawa | Sqw | I03 + B01 | |

[*]References for the spectrum and the previous Bus-DeMeo taxonomy: B88, Bell et al. (1988); C95, Clark et al. (1995); B01, Binzel et al. (2001); B02, Bus & Binzel (2002a); I03, Ishiguro et al. (2003); C04a, Clark et al. (2004a); C04b, Clark et al. (2004b); B05, Barucci et al. (2005); Ha06, Hardersen et al. (2006); S08, Sunshine et al. (2008); Ha11, Hardersen et al. (2011); C09, Clark et al. (2009); D09, DeMeo et al. (2009); Y10, Yang & Jewitt (2010); T12, Takir & Emery (2012); V14, Vernazza et al. (2014); R16, Reddy & Sanchez (2016); SMASS, Database of The MIT-UH-IRTF Joint Campaign for NEO Spectral Reconnaissance[5].